# Translation of Enterprise Architecture Concept to Facilitate Digital Transformation Initiatives in Vietnam: Processes, Mechanisms and Impacts

*Completed Research Paper*

Duong Dang
University of Vaasa
Wolffintie 32, 65200 Vaasa
duong.dang@uwasa.fi

Quang "Neo" Bui
Rochester Institute of Technology
Rochester NY 14623
qnbbbu@rit.edu

## Abstract

*Governments around the world have increasingly adopted digital transformation (DT) initiatives to increase their strategic competitiveness in the global market. To support successful DT, governments have to introduce new governance logics and revise IT strategies to facilitate DT initiatives. In this study, we report a case study of how Enterprise Architecture (EA) concepts were introduced and translated into practices in Vietnamese government agencies over a span of 15 years. This translation process has enabled EA concepts to facilitate various DT initiatives such as e-government, digitalization, to name a few. Our findings suggest two mechanisms in the translation process: a theorization mechanism to generalize local practices into field-level abstract concepts, making them easier to spread, while a contextualization mechanism unpacks these concepts into practical, adaptable approaches, aligning EA with adopters' priorities and increasing its chances of dissemination. Furthermore, our findings illustrate how translation happened when the initial concepts are ambiguous and not-well-understood by adopters. In this situation, there is a need for widespread experiments and sense-making among pioneers before field- and organizational-level translation can occur.*

# Translation of Enterprise Architecture Concept to Facilitate Digital Transformation Initiatives in Vietnam: Processes, Mechanisms and Impacts

*Completed Research Paper*

## Introduction

In recent years, countries around the world have aggressively adopted digital transformation (DT) initiatives to strategically increase their competitiveness in the global market (European Center for Digital Competitiveness 2021). Despite the promise of significant benefits, the adoption of digital technologies such as machine learning (ML), artificial intelligence (AI), Internet of Things (IoT), or cloud computing tend to be disruptive, often require a comprehensive transformation of existing processes, IT architectures, human resources, and business strategies (Eden et al. 2019; Vial 2019; Wimelius et al. 2021). Subsequently, numerous public organizations have encountered challenges in their DT initiatives, stemming from incompatibilities with legacy systems, organizational complexities, and bureaucratic red tapes (Irani et al. 2023; Zaman et al. 2022). One thing is clear, to make DT initiatives work, public organizations need a comprehensive revision of not only their IT assets but also of their IT strategies.

To this end, various organizations have revised their enterprise architecture (EA) to accommodate their DT initiatives (Dang & Pekkola, 2023; Dang & Vartiainen, 2022; Do et al., 2023; Thai et al., 2022). As a strategic tool and an IT governance mechanism, enterprise architecture enables the logical organization of IT infrastructure and capabilities to support business strategies (Ross et al. 2006; Bui 2017). Within the US public sector, EA has been a key initiative to reduce IT complexities and costs (GAO 2002). In the wake of DT initiatives, organizations have evolved their EA in support of their transformation journey. Yet, this evolution process is not well-understood, leaving a research opportunity to further explore how organizations revise their current IT governance framework to support DT initiatives.

In this paper, we address this research gap through a case study of EA evolution in Vietnamese government agencies to support the country's DT initiatives. The case of Vietnam is particularly interesting because (1) it offers an opportunity to observe how the EA concepts were adopted from the US public sector and then translated into Vietnamese government agencies; and (2) it offers an opportunity to scrutinize the DT process in Vietnam, a country ranked as the number one "digital riser" in the East Asia and Pacific region (European Center for Digital Competitiveness 2021). Our overarching research question is: *how do the EA concepts get adopted and evolve over time to facilitate DT initiatives in Vietnamese government agencies?*

Our paper makes two contributions to the increasing number of studies on translation and innovation adoption (Nielsen et al. 2013; Nielsen et al. 2020). First, our findings suggest that the translation of EA concepts involves two key mechanisms: a *theorization* mechanism that generalizes local EA practices to field-level abstract concepts that are easy to diffuse; and a *contextualization* mechanism that unpacks field-level abstract concepts into situated practices that are increased the practicality and compatibility of EA concepts to potential adopters. Furthermore, while theorization legitimizes EA concepts and attracts early adopters, contextualization aligns EA with the priorities of the majority adopters (e.g., DT initiatives) and increases its chance to spread. Second, our study provides a contrasting view to prior studies that documented a translation process of clearly-defined and understood innovation concepts (e.g., Leadership Pipeline research (Nielsen et al. 2020), mobile IT in Danish homecare agencies (Nielsen et al. 2013). Instead, our study showed a translation process of an ambiguous innovation concept that needs experiment, sense-making, and field-level translations. As a result, our findings provide a more detailed translation process.

The rest of the paper is as follows. We first present the theoretical background on translation theory which is the basis of our empirical inquiry. Next, we present our methods and findings. We conclude with a discussion on the theoretical and managerial contributions.

## Theoretical Background

We use translation theory (Czarniawska and Sevón 1996; Nielsen et al. 2013) to guide our empirical inquiry and theoretical development. The translation theory in fitting here because in our empirical data, we observe the "travel" of EA concepts from the US public sector into Vietnamese government agencies (Czarniawska and Joerges 1996) and the translation of EA concepts between the field- and organizational level (Nielsen et al. 2013).

Translation is a process of "creating convergences and homologies by relating things that were previously different" (Callon 1981, p. 211). It is a complex process of negotiation during which meanings, claims and interests change and gain ground. Translation has several meanings (Wæraas and Nielsen 2016): (1) a political meaning, referring to the "pursuit of interests or specific interpretations, frequently involving acts of persuasion, power plays and strategic maneuvers" (Wæraas and Nielsen 2016, p. 237); (2) a geometric meaning, referring to the "mobilization of human and nonhuman resources 'in different directions', the result of which is 'a slow movement from one place to another'"(Wæraas and Nielsen 2016, p. 237; cited Latour 1987, p. 117); and (3) a semiotic meaning, which "concerns the transformation of meaning that occurs during the movement of the object in question" (Wæraas and Nielsen 2016, p. 237).

Within the innovation studies, prior studies have established that the diffusion process can be conceptualized as a translation process of innovative ideas rather than the actual diffusion of innovative practices (Nielsen et al. 2013). Prior studies have examined the "diffusion as a translation process" idea from three different levels (see Table 1). First, translation can happen at the micro-level in which organizations take field-level concepts and translate them into situated practices. Studies within this approach focus on mechanisms that enable the movement of innovative ideas into local contexts and institutionalize the new practices (Niemimaa and Niemimaa, 2017; Wagner et al., 2018). Second, translation can happen at the meso-level in which innovative ideas travel across organizations within a community. Studies such as Waardenburg et al. (2022) or Anderson et al. (2023) examine the role of knowledge brokers facilitating the translation, as well as specify the phases in the translation process. Finally, translation can happen at the dynamics of micro- and meso-level in which innovative ideas travel across organizations within a community, and such ideas are modified and translated into situated practices at the organizational level (Nielsen et al. 2013; Nielsen et al. 2020).

It is the third approach that our paper aims to contribute to. Due to the complexity of this phenomenon, studies within this third approach are scarce, leaving several questions unanswered. For instance, how do field-level translation differ from organizational-level translation? Or what are the mechanisms at play throughout the translation process? In the next section, we describe our method and empirical inquiry in attempting to answer those questions.

| Level | Translation mechanisms |
|---|---|
| Micro-level: Within organizations that are interested in an innovation | - Three translation mechanisms: (1) translating global to local, (2) disrupting and reconstructing local non-canonical practices, and (3) reconstructing and enacting local canonical practices (Niemimaa and Niemimaa 2017).<br>- Three translation mechanisms that support local work practices by connecting them to the broader public ideas upon which the practices are based: doctrine, identity, and norms (Wagner et al. 2018). |
| Meso-level: Across organizations within a community | - Translation facilitated by knowledge brokers who championed the knowledge transfer and held three consecutive roles -- messenger, translator, and curator (Waardenburg et al. 2022)<br>- Translation of regulations into policies relies on pragmatic reasoning schemas, which can lean towards permissions or obligations. Balancing these schemas by translation guardrails is crucial to translate regulations into effective policies that avoid being excessively lax or restrictive (Anderson et al. 2023). |
| | - The reciprocal translation of smart city concepts and bureaucracy reveals three translation mechanism: comprehension, filing, and standardization for bureaucracy (Khodachek et al. 2022). |
| Micro-and-macro level: Within and across organizations within a community | - The mechanism was that the ideas about mobile IT usage has been articulated and legitimized in the Danish home care field. It then materialized in the practices of individual home care agencies (Nielsen et al. 2013).<br>- The mechanism involved translating the U.S.-born Leadership Pipeline concept from the private sector to the Danish public sector, primarily through field-level modifications and the publication of a book titled "Leadership Pipeline in the Public Sector," resulting in the spread and application of the Leadership Pipeline concept within Danish organizations (Nielsen et al. 2020). |
| Table 1. Summary of Translation Studies | |

# Methods

## Data Collection

To examine the translation of EA concepts in support of DT initiatives, we conducted a case study of EA translation in the Vietnamese government agencies from 2005 to 2022. We chose Vietnam as our empirical setting for both theoretical and practical reasons. Theoretically, EA was a part of the World Bank-funded project (thereafter the WB Project) for the Vietnam government for ICT development. It thus offers an opportunity to observe a field-to-field translation process in which the EA concepts travel from outside i.e. the US public sector and then get translated into Vietnamese government agencies. Also, From the managerial perspective, Vietnam is ranked as the number one "digital riser" in the East Asia and Pacific region (European Center for Digital Competitiveness 2021). Hence, the case will offer valuable insights to other developing countries that are also adopting DT initiatives. In addition, one of the authors was actively involved in the DT initiatives in Vietnamese governmental agencies. As a result, we were allowed unrestricted access to case materials, allowing us to have a rich conceptualization of the case.

Table 2 provides a summary of the collected data. The data collection included interviews, observations, archival data, and exchanges following up emails. We first gained an understanding of the phenomena by being involved in several observations and studying archival data. In particular, we participated in seven sessions for in total of 21 hours, including the following provinces Ho Chi Minh City (HCM), Ha Tinh (HAT), Tien Giang (TIE), Can Tho (CAN), Da Nang (DNG), Quang Ninh (QUA), Lao Cai (LAO). Then, the interviews were conducted from June 2015 to August 2015, and from July 2016 to August 2016. We used semi-structured interviews with open-ended questions (Myers 2019), focusing on EA concept development in each case: problems, process, pressure, institutionalization, and legitimacy. At the field level, we interviewed a total of 13 MIC officials who were in charge of IT applications in state agencies within the country, including the CIO, project managers, IT architects, and officials. At the organization level, we interviewed provinces that had EA programs or EA projects, including HAN, HAT, and QUA. A similar procedure to the field level was applied to provinces. We also interviewed one consultant who worked as consultancy services for the World Bank-funded project for Vietnam. Moreover, we collected archival data simultaneously with interviews, those archival data include the internal project documents such as meeting notes, drafts, proposals, plans, and policies. In total, we collected more than 300 documents of about 6000 pages. Finally, we used email to follow up or discuss the phenomena. For example, before the trip to DNG, we emailed the DNG department of ICT to discuss their EA and asked them to provide materials in order to understand DNG EA. Doing so, helped us have an overview picture before an observation session took place. In total, we had 145 emails.

| Data Type | Source | Quantity |
|---|---|---|
| Interviews | World Bank consultants<br>Ministry officials | 1<br>13 |

|  | Municipal officials | 27 |
|---|---|---|
| Archival data | Office visits, project presentations, and workshops on DT initiatives at the municipal agencies | 7 sessions, a total of 21 hours |
| Observations | Internal project documents (e.g., meeting reports, drafts, strategic plans) Public policies | Total of 300+ documents (6,000+ pages) |
| Email exchanges | Follow-up emails to clarify concepts or discuss phenomena | 145 emails |
| Table 2. Summary of Collected Data ||| 

*Data Analysis*

We follow a case study approach (Yin 2009) to analyze our data. First, interviews were transcribed and key points from the interviews and secondary data were summarized. Doing so helped us to significantly reduce the number of pages to a manageable level, allowing us to identify key facts and points. Then, we transferred the data to NVivo to assist us in data analysis. We cross-checked interviews with secondary data. For observation and workshop data, we cross-checked or verified through materials and emails exchanged between the researcher and officials in charge in provinces. It is worth noting that the original data are available in both English and Vietnamese. For instance, all materials related to World Bank-funded projects were in English and in Vietnamese (EAs in HAN, MIC, and DNG), while materials for EA projects or programs in other organizations were in Vietnamese. In the latter case, we did not translate all Vietnamese materials into English; only quotations were translated in the writing-up phase. Also, we followed Myers' (2019, p. 45) suggestion that "a social researcher must already speak the same language as the people being studied […]." As a result, interviews were conducted in Vietnamese although questions were available in both English and Vietnamese.

After the data were processed, we built a case description to holistically make sense of the events in the case. Then, following (Miles and Huberman 1994), we built a conceptual map of the events to identify their relationships (Figure 1). Process tracing techniques were used to identify missing links in the map. Finally, we engaged in iterative rounds of theorization to identify the mechanisms that underlie the translation process in our case. We frequently go back to the literature to connect our findings with prior studies and to make sense of our data. It was this process that allowed us to identify the translation process as the theoretical basis for our study. Once we reached theoretical saturation and a complete understanding of the case, we stopped our analysis. Throughout the process, key quotations from the interviews were selected. For quotations were in Vietnamese, two authors were involved in the translation from the original Vietnamese text into English. The first author translated independently and put the Vietnamese text along with the English. The other author then cross-checked. If conflicts arose the two authors discussed to reach a consensus agreement about the matter.

# Findings

In this section, we report the findings on the translation of EA concepts in Vietnamese government agencies to facilitate their DT initiatives. Figure 1 shows a conceptual map of the events that happened in the case. We roughly organize them into three phases based on 1) our theoretical sense-making of the data and 2) theoretical terms used in translation studies and institutionalization theory (Nielsen et al. 2013; Nielsen et al. 2020; Tolbert and Zucker 1996).

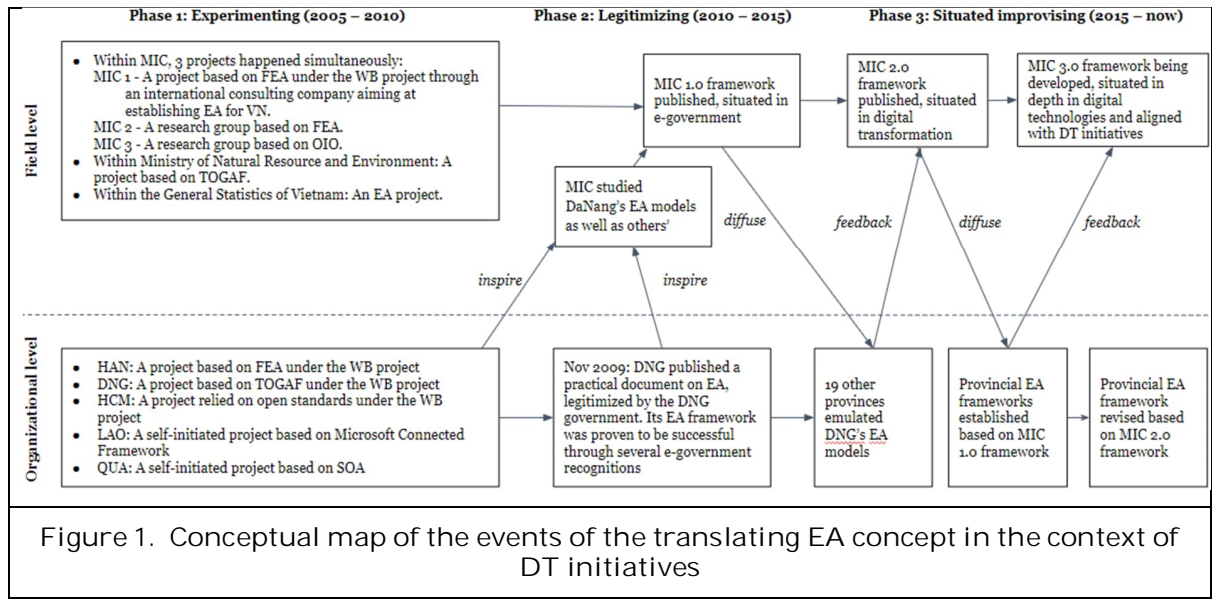

Figure 1. Conceptual map of the events of the translating EA concept in the context of DT initiatives

*Phase 1: Experimenting (2005 – 2010)*

The first phase can be described as an "experimenting" phase. During this phase which lasted from 2005 to 2010, the EA concepts traveled from the US and were introduced into the Vietnamese public sector through the work of a knowledge broker. However, these EA concepts were too ambiguous, forcing pioneers to sense-making EA concepts to understand what they mean. Then, various adopters at both the field level and organizational level attempted to start their own EA frameworks. Hence, this phase was characterized by the experiments of these adopters with the newly-introduced EA concepts.

Introducing EA Concepts into the Vietnamese Public Sector

The EA concepts were first brought to Vietnam through the WB Project sponsored by the World Bank in 2005. One of the aims of the project was to promote ICT development in Vietnam at both the national level and the municipal level i.e. provincial level. Among its lofty goals, the project aimed to develop "a national ICT architecture and interoperability framework for e-government" (The WB Project Appraisal Document 2005; The WB Project Credit Agreement 2005, p. 16). The original development of ICT architecture and interoperability framework was centered around three frameworks: the Dublin Core, the Federal Enterprise Architecture of the US and e-Government Interoperability Framework (eGIF) of the UK. It also referred to the European Interoperability Framework for pan-European e-government services (The WB Project Appraisal Document 2005).

As the WB Project unfolded, five sub-projects related to EA were initiated. At the field level, EA concepts were adopted by the Ministry of Information and Communication (MIC) and the General Statistics Office of Vietnam (GSO). For instance, MIC was charged with the task to develop "a national-level IT architecture and interoperability framework to establish common standards across government" (The WB Project Credit Agreement 2005). On the other hand, three sub-projects related to EA were initiated at the organizational level (i.e., in municipalities or provinces). The sub-projects were led by agencies in HAN, HCM, and DNG, the three biggest cities in Vietnam. At this level, EA concepts were part of a program to develop "ICT strategy, policies, architecture and interoperability framework; and review and upgrade of ICT infrastructure, equipment and systems; […]" (The WB Project Credit Agreement 2005, p. 17).

It is worth noting that unlike other concepts in prior translation studies, EA concepts when introduced into Vietnam were highly ambiguous, and there was no concrete definition of EA concepts in any WB Project documents. Instead, several terms were used, such as "IT architecture" (The WB Project Appraisal Document 2005; The WB Project Integrated Datasheet 2005), "ICT architecture" (The WB Project Credit Agreement 2006), "e-Government architecture", "IT architecture" or "shared IT architecture" (The WB Procurement Plan 2007), and "e-government Enterprise Architecture" or "IT architecture" (The WB

Restructure of the Project 2010). Within the subprojects, various terms were used. For instance, terms such as "Enterprise Architecture", "national IT architecture", and "E-Government Enterprise Architecture" were used in field-level sub-projects while terms such as "City-level IT architecture", "IT architecture", "e-Govt Enterprise Architecture", and "EA Master Plan" were used in HAN and DNG subjects (The WB Completion Report 2014). The proliferation of terms only ended when the definition of EA was first introduced in 2014. The definition was drawn from Lankhorst (2005)'s definition and EA was defined as a "high-level strategic technique to help senior managers achieve business and organizational change; a coherent set/whole of principles, methods and models used in the design and realization of an enterprise's organizational structure, business processes, information systems and infrastructure." (The WB Completion Report 2014, p. 31)

The ambiguity in EA concepts could be due to the lack of a universal definition of EA around the world. One MIC official who was involved in writing the Request for Proposal and Terms of Reference (TOR) for the WB project said:

> "EA concept itself does not yet have a unified understanding between different countries, but EA is being understood based on an imaginative perception about it in the sense that it [EA] is something that serves as a guide to building an effective and efficient e-government. This is the perception and common sense of all countries about EA. Based on the specific conditions of each country, they have different approaches on how to build EA for e-Government effectively […]. Because of the different approaches, the output results are different and appropriate to the context of each country."

As a result, in this period, early adopters of EA concepts had to interpret EA concepts on their own and experimented with various frameworks. Next, we describe these experiments at the field- and organizational-level.

Experimenting EA Concepts at the Field-level

At the field level, with the influence of the ongoing WB project at that time, the government wanted to establish a national IT architecture. In 2007, a government decree charged the MIC with the responsibility to "preside over building the national architecture standard for information systems" (Government Decree No. 64 2007). However, in this decree, the government did not state what this national IT architecture standard for information systems means, leaving the interpretation and implementation to MIC. Not surprisingly, it created confusions and even frustration among MIC officials. Several of them commented:

> "[…] the Decree has only one sentence but do not mention the content [of what EA is] making it impossible to do [EA]" (MIC official)

> "In the time of this decree [No. 64, 2007], the WB project was ongoing and had EA sub-projects (in MIC, HAN, DNG, and HCM), so they included that clause. The ironic thing is that 10 years later, the concept of EA is still not clearly defined" (MIC, Deputy Head of Strategy Unit AITA).

To conform with the decree, at MIC, the national IT architecture and other EA concepts were built based on the US' Federal Enterprise Architecture framework through a consulting company (III- a company from Taiwan). At the same time, they established two research teams: One focused on studying FEA in-depth, and the other focused on the Danish government enterprise architecture framework (OIO) of Denmark. Moreover, a national research project on EA was also established under the leadership of MIC. The result was very limited in this period. MIC was not able to legitimize EA concepts, and they stuck to proposing a framework that would not be applicable to state agencies within the country. The lead consultant, who was one of the key persons in charge of consultancy services for both MIC and HAN's EA projects, commented:

> "Enterprise architecture was born to serve the [private] organization as its meaning of "enterprise" indicates consistency from top to bottom. EA for the public sector is more complex, decentralized, and deauthorized [compared to the private sector] […] enterprise architecture must reflect not only the architectures, but it also business, culture, and customs/traditions […], in Vietnam, there are several such characteristics: the admin with a hierarchy of three-level decentralized mechanisms, and decentralized budgets. This leads to decentralization of power and budget, making EA difficult to work [in practice] because others may not want to accept it [result or task from others]."

Besides MIC, EA concepts were also adopted by other ministry-level agencies such as the General Statistics Office (GSO) of Vietnam or the Ministry of Natural Resource and Environment. However, similar to MIC, they also suffered a low success rate.

Experimenting EA concepts at the Organizational Level

As a result of the WB project, three provinces (HAN, HCM, and DNG) have implemented WB subprojects. Each subproject had its own Project Implementation Unit (PIU) to manage and facilitate the subproject in their province. In other words, each subproject had complete control over how implementation could unfold. Subsequently, each province experimented EA concepts in its own way, and the following describes their efforts:

*Experiments at HAN*: Institute for Information Industry of Taiwan and IVTECH of Vietnam were two companies providing consulting services to develop strategy, roadmap and IT architecture and standards for Hanoi. Originally, EA concepts were based on FEA and TOGAF. FEA was used as a reference to build business and service reference architecture, application architecture, data architecture, and technology architecture, while both FEA and TOGAF were used for the maintenance of EA. However, in the final draft version of the project, the one that will be applicable to all agencies in HAN if approved, it was solely based on FEA (Draft of HAN EA 2014). Unfortunately, the draft was not approved by the governor. One consultant explained: One of the main reasons because the lack of awareness of EA as one consultant explained:

> "When I worked for HAN [EA consultancy services], people didn't understand when I presented and reported. When I presented and reported to [the HAN leader who oversees HAN's EA project], he kept asking why I put such a model in [...] when presenting [EA], people sit and listen and watch but they don't understand anything."

Another reason why the EA draft could not be realized was because of the complexity in public agencies. One of the members in HAN's EA team explained:

> "[...] HAN has approximately 6000 public services, owned by different departments, districts, and communes. It was very difficult to design business architecture that was covered and agreed upon by owners. In practice, it doesn't work [for EA that covers all of them], and that is why it's blocked."

As a result, after five years and working with two consulting companies, HAN could not legitimize EA concepts and produced a formal EA framework.

*Experiments at HCM*: HCM was first proposed as part of the WB subprojects with a budget of $9.47 USD million. However, this budget was eventually canceled; and HCM instead used its own budget to carry out the activities proposed under the project. In particular, they developed a framework for e-government called "HCM E-Government Framework" that aimed to consolidate all services and applications in state agencies belonging to HCM to this only platform based on open source software and technologies. In particular, the framework includes infrastructure architecture, portal architecture, core software architecture, application architecture, official architecture, and citizen and organizational architecture. Commented on the approach chosen by HCM, one official at MIC stated:

> "HCM's EA has a better approach in comparison to HAN's EA approach [by contextualizing EA that is suitable to HCM] [...] HAN chooses an approach for EA as a strategy with as-is and to-be and make a recommendation to agencies to follow, while HCM translates EA as a technical framework and HCM proposed a framework to connect information systems together, align with their current information systems for e-government and force agencies to connect to its framework. That makes it work and agencies see the results immediately [...]."

Due to this approach that focused more on technical implementations of EA concepts for e-government, the EA concepts were successfully adopted in HCM to support its e-government portal.

*Experiments at DNG*: EA was one of the main tasks in DNG's WB sub-project; and DNG chose to follow TOGAF when implementing EA. In its EA framework, DNG coupled EA concepts with public services and e-government. Specifically, EA was built to couple with Service-Oriented Architecture, among others. It promoted loose coupling, reuse, and interoperability between systems. Also, EA helped to balance between excessive and no controls over IT systems, allowing more flexibility and adaptability in IT systems to e-government requirements. As a result, DN enjoyed great success with EA as it was the only province with an EA in place by November 2009 to guide the city's e-government development. The Head of the DNG Department of ICT commented on success factors for EA in DNG as follows:

> "Leadership's determination [to support EA project in DNG]; The companion of all districts within DNG; the absolute coordination with the Department of Home Affairs [that is in charge of public services and administrative duties]" (Note taken from the Observation of 2014).

Experiments at DNG were considered as a success given that at that time all the WB-funded EA projects in other organizations had no clear results or made little progress, while in contrast DNG was successfully integrated its EA in its e-government framework and became a leading province within the country in IT applications and services.

Apart from those key EA sub-projects prompted from the WB Project, there were several provinces and ministries that also developed EA on their own. For example, LAO developed EA in 2008 based on the Microsoft Connected Framework (MCF); QUA developed its own EA based on SOA. The reason the EA concepts also spread into other provinces and ministries was that the EA concept was discussed at the CIO Council of Provinces and the CIO Council of Ministries, as well as the National Committee on IT Applications in state agencies (chaired by the Deputy Prime Minister, whose members are ministers of government ministries). Unfortunately, DNG was the one with the most success while others either failed or did not get much headlight from the public.

Summary

Overall, the first phase witnessed many experiments from pioneers in the Vietnamese public sector to adopt EA concepts. The process, mechanism, and impact of this phase can be summarized as follows.

- *Process*: the 2005 WB Project introduced EA concepts to Vietnamese government agencies, triggering five sub-projects to adopt EA (i.e., micro translating EA concepts into organizations)
- *Mechanism*: a sandbox approach in which various pioneers simultaneously adopted EA concepts; most failed and a few succeeded
- *Impact*: various experiments allowed pioneers to try different ways to translate EA concepts into practice as well as drew lessons on what made EA work

### Phase 2: Legitimizing (2010 – 2015)

The second phase is characterized by efforts to legitimize EA concepts to the broader public sector. From 2010 to 2015, the works of the most successful pioneers in the first phase were traveled up to field-level organizations. Here, EA practices were generalized into abstract frameworks that allow them to be understood by others. We capture the events in this period regarding 1) the success of experiments at organizational-level, 2) the pressure to legitimize EA at the field level, and 3) the publication of an EA framework at the field level.

Success of Organizational-level Experiments

At the organizational level, on July 14, 2010, the EA framework for IT applications for DNG was approved by the DNG mayor (DNG Decision No. 5258 2010), legitimizing EA concepts and practices. At the time, DNG defined EA as "a methodological approach that provides the ability to analyze multidimensional aspects of an organization in an effective way to combine the strategic level of organization with the operational requirements of the organization" (DNG Decision No. 5258 2010). The framework was based on TOGAF. At the same time, DNG e-government which embraced the EA framework was ranked No. 1 in the country (out of 63 provinces) and received several awards such as eAsia Award 2013, WeGo 2014 for bridging the digital divide, and FutureGov 2011 on e-government development. These recognitions reinforced the legitimacy of EA practices at DNG. Subsequently, several provinces in the country emulated DNG's approach toward their EA development to digitalize government. By 2015, the EA concepts were transferred from DNG to 19 other provinces in the country.

In a similar vein, HCM also developed and integrated its own EA into its e-government platform. This proved successful as HCM was among the top ten rankings in e-government within the country. The HCM approach also attracted others and HCM was willing to transfer its model. For example, they organized a large hands-on workshop from September 29th to October 4th, 2014 dedicated to transferring EA. In the workshop, 26 provinces attended, including DNG. It is worth noting that HCM was one of the largest provincial-level cities in the country with about 8 million people. They had the capability, resources, infrastructure, and knowledge to develop and maintain their EA. This was in contrast to most of the other provinces in the country with a population of around 1 million and limited resources. The head of the IT section of HCM's ICT department commented:

> "HCM's model is very good, but to apply it requires a lot of resources including human, financial, and local policies and commitments. That is why not many provinces can apply the HCM model in my view."

### Field-level Pressure to Legitimize EA

At the field level, there were several tries to develop and translate EA into use. For example, the Ministry of Natural Resources and Environment established its own EA framework in 2014 based on TOGAF. Others included the Ministry of Finance and the Ministry of Health. However, these efforts did not attract wide attention from other government agencies. In the meantime, MIC, which was the agency in charge of EA development in the country, did not make substantial progress regardless of several attempts such as research groups or EA development projects. As the WB Project progressed, it built pressure on provinces that eventually turned into pressure for MIC to provide field-level guidance and standards. Under the mounting pressures, the Deputy Prime Minister decided that MIC had to study DNG's EA and if it worked, extend DNG model to the entire country (Government Decision No. 313 2014).

Within MIC, the Authority of Information Technology Applications (AITA) unit was in charge of studying the DNG model. Subsequently, AITA organized a field trip to DNG in November, 2014. During the trip, there was a conflict between MIC and DNG regarding what should be included in DNG model: whether or not DNG model includes enterprise architectures, infrastructure for IS, e-government framework (for public services and backend systems), and resources, policies needed for implementation of DNG model. This conflict was reflected by an official who was a part of the AITA team:

> "There was a difference in understanding EA between DNG and MIC: they were more focused on operational while we were more focused on strategic. We had 30 meetings, and email exchanges before the field trip and all discussions focused on nine areas: How to seek a common understanding of DNG model; EA; ICT infrastructure of DNG; e-government platform for public services and backend IT systems; resources, and policies needed for the DNG model; interoperability capability of DNG; effectiveness; Conformity between DNG model and current regulations and standards in the field of information and communications; and security capability"

At the same time, the national committee on IT applications in state agencies organized two events: (1) a workshop on digitalization government on August 14th, 2014 for about 30 provinces in the Southern part of the country and EA was one item in the workshop agenda; and (2) a similar workshop on August 21st, 2014 for about 30 provinces in the Northern part of the country. In these events, DNG was presented and discussed. The lessons learned in these workshops became the basis for many subsequent EA frameworks implemented by provinces.

### Publication of a Field-level EA Framework

Field-level events mounted up to concrete actions, and on April 21st, 2015, the Enterprise Architecture framework for Vietnamese e-government version 1.0 (EA 1.0) was released, more than five years after the first organizational-level EA was legitimized. EA 1.0 was based on provinces like Danang, Gartner's e-government architecture, and FEA. One of the officials at MIC commented on EA 1.0:

> "The framework issued by Vietnam [EA 1.0] has only achieved certain goals. The framework has not included many architectures, such as business architecture, data architecture, and information architecture. [...] but I think it only shows the technical architecture or something like that, the framework is just that and for the other parts, I share with you that we don't have a methodology when it comes to establishing an EA [...]. In my opinion, we often have a prejudice that we use the American framework [FEA] as our standard, so now everything we do [regarding EA] is compared

> to it. But compared to it, it turns out to be difficult [...] I believe other EAs in the world probably also have that prejudice."

At first, EA 1.0 was very abstract and difficult to apply. One of the key persons who led a group behind EA 1.0 posed:

> "MIC built EA to provide some content, some step-by-step guidance and a few ministerial and provincial-level administrative frameworks at the most general and abstract level. To my view with an academic perspective, it [EA 1.0] cannot be considered an enterprise architecture. Instead, this is a step-by-step framework to build an architecture that presents two models at provincial and ministry levels in the most basic and general way [...] [EA 1.0 released] due to pressure, MIC had research on EA, along with the WB subproject and the results of all of them encountered many difficulties, so MIC came up with EA 1.0 [...]. Leave it for provinces and ministries to develop and implement their detailed EA as EA 1.0 only provides general guidance."

Subsequently, the ambiguity and lack of details in MIC 1.0 caused confusion and put pressure on provinces to interpret it themselves. These issues motivated MIC to subsequently evolve its framework, captured in the next phase.

Summary

The second phase witnessed some successful experiments at the organizational level, which prompted field-level adoption to release the first national EA framework. This framework heavily leaned on e-government which was the primary focus at the time. The process, mechanism, and impact of this phase can be summarized as follows.

- *Process*: success of pioneers inspires MIC to study success stories and came up with the first national EA framework, focusing mostly on e-government applications
- *Mechanism*: a theorization process in which lessons learned from DNG and other success stories get formalized into an EA framework
- *Impact*: 19 provinces emulated DNG's approach, and the first national EA framework was published

*Phase 3: Situated improvising (2015 – 2022)*

The third phase is characterized by efforts from MIC to improvise and evolve EA frameworks to support various digital transformation initiatives carried out in provinces. From 2015 until now, MIC has updated the EA framework to version 2.0 that focused on digital transformation, and version 3.0 that focused in-depth on digital technologies and aligned with DT initiatives. Below, we describe 1) the concerns surrounding EA 1.0, and 2) the development of EA 2.0 and 3.0

Concerns Regarding EA 1.0

After EA 1.0 was realized, MIC was in charge of guidance in the development of EA across the country. The process of establishing an EA in an organization was that each organization (province) drafted its EA, and sent it to the MIC for comments before approval. Also, MIC dedicated a unit that focuses on consultancy EA development in organizations. EA 1.0 was general enough so that it fit with every EA that was proposed by provinces which also made it easy to diffuse EA concepts to interested provinces. However, this also created a dilemma for some provinces as commented by an official for QUA province:

> "[...] MIC's EA 1.0 is more conceptual, meaning it has this and that but people [at the organizational level] don't know what to do. There are no criteria corresponding to that concept, so there is no roadmap and resources to do it [...] the challenge is that with or without it [EA 1.0], the provinces will still do it. EA 1.0 hinders the development of EA in provinces as you can establish your EA in whatever you want and it still complies with EA 1.0. In other words, MIC EA 1.0 would make things difficult for the local agencies [...] In the current version, MIC EA 1.0 leaves the provinces to fiddle with it, not every province spends several hundred billion [Vietnamese Dong] to make investments in EA programs. You may have EA [as a document], but no one will carry it out because they don't have the resources to implement it."

In other words, provinces referred to EA 1.0 as a means to develop and legitimize their EA. But in reality, they translated EA on their own. A consultant commented on why EA 1.0 was released even though it got criticism from others:

> "[…] coming up with the EA 1.0 unifies awareness and methods, but [EA 1.0] does not specify anything. How to implement it in organizations is a serious question both at national and provincial levels […] when it comes to EA there are two approaches: do it methodically [in a radical way] or to change gradually […] [and MIC 1.0 chose the latter way]."

Nevertheless, EA was quickly translated to organizations across the country. By 2018, 61 out of 63 provinces and 19 out of 22 ministries had their legitimized EA framework built on EA 1.0 and approval by the local governments. The Head of Planning and Investment at AITA explained why EA was rapidly adopted by organizations:

> "Why is it happening? [there are several factors, such as] political pressure, a lot of investment funds, projects, and research projects so there is pressure to produce something, and this is internal pressure. The external pressure is that if the other countries do it [development EA], we should do it too. Many countries have done it so we should follow the trend. Money and resources have been invested so there must be results."

### Publication of EA 2.0 and 3.0

On December 31, 2019, the MIC issued the Vietnam e-Government Architecture Framework, version 2.0 (EA 2.0) (MIC Decision No. 2323 2019). This is a guiding document on building ministerial-level and provincial-level EAs. In comparison to MIC EA 1.0, the 2.0 version added a lot of more detailed guidelines and situated around digital transformation issues, something that was of high priority for provinces. Specifically, EA 2.0 included (1) the principles for building EA; (2) the concept of EA, (3) Ministerial-level EA recommendations, Provincial EA recommendations; (4) Adding reference models: Business Architecture; Data architecture; Application architecture; Technology architecture; Information security architecture; (5) Align with current national ISs and database and technology development trends such as Cloud Computing, Big Data, and AI. Moreover, EA 2.0 was positioned to better fit the priorities of adopters, as well as to align with DT initiatives and governmental digital strategy, as stated in the Government Resolution:

> "MIC should establish EA 2.0 for the period of 2019-2025 to make sure EA 2.0 […] to integrate ministerial-level e-Government, provincial-level e-Government, national information systems and national databases, ensuring compatibility with the context of the fourth Industrial Revolution and the trend of e-Government development in the world". (Government Resolution No. 17 2019)

Subsequently, Provinces and Ministries started to upgrade their EA compliance to MIC EA 2.0. For instance, HAN—one of the pioneers that failed to translate EA into actual use in the Phase #1 was able to adopted MIC's EA 2.0 and released its own EA frameworks in 2021 (HAN Decision No. 4097 2021).

By the end of our data collection, MIC was in the progress of releasing EA 3.0. It was reviewed and updated to have significant ties with national DT initiatives. For instance, in the Directive No. 05, 2023 of the Vietnam Government, it stated, "Review and update EA 2.0 […] to align with national DT initiatives, national databases, […] establish digital platforms, EA 3.0 to serve the national digital transformation period 2022 - 2025, with a vision to 2030". While the success of EA 3.0 remains to be seen, evidence from our conversations with different stakeholders shows a promising future for EA in Vietnamese government agencies, particularly in its digital transformation process.

### Summary

The third phase witnessed many improvising from MIC to better fit EA frameworks to the priorities and needs of adopters. The process, mechanism, and impact of this phase can be summarized as follows.

- *Process*: EA 1.0 was too abstract leading to confusion among adopters; EA 2.0 and 3.0 were subsequently released with more details and situated practices that relevant to adopters (e.g., digital transformation, digital technologies

- *Mechanism*: a contextualization process in which abstract concepts in EA 1.0 were expanded and contextualized in issues that are relevant to majority of the adopters
- *Impact*: increase practicality and compatibility of EA concepts; many provinces and ministries upgraded their EA framework to comply with EA 2.0 and 3.0

## Discussions and Conclusions

In this paper, we examine how EA concepts get adopted and evolved over time to facilitate DT initiatives in Vietnamese government agencies. Through a case study of EA adoption from 2010 to 2022 across Vietnamese provinces and ministries, we show a dynamic evolution of EA concepts: (1) EA was first introduced through a knowledge broker, the World Bank, followed by a period of rapid experiments by several pioneer governmental agencies; (2) success from organizational-level agencies inspired field-level theorization and formalization of EA concepts, making them more attractive to other provinces; and (3) feedback on early EA frameworks triggered revisions of field-level EA frameworks to better fit with current DT initiatives prioritized by agencies. Table 3 provides a summary of our findings.

Our study offers two contributions to translation theory and innovation diffusion studies (Nielsen et al. 2013; Nielsen et al. 2020). First, our findings suggest that the translation of EA concepts involves two key mechanisms. First, a *theorization* mechanism that generalizes local EA practices to field-level abstract concepts that are easy to diffuse. This is evident in the second phase when elements from the successful EA framework in DNG were picked up by MIC and formalized into EA 1.0. By publishing EA 1.0, other provinces were able to get the necessary guidance to jumpstart their own EA adoption. Second, there is a *contextualization* mechanism in the translation process that unpacks field-level abstract concepts into situated practices that increase the practicality and compatibility of EA concepts to potential adopters. This is evident in the third phase when MIC received feedback on the lack of details in EA 1.0 and published EA 2.0 and 3.0 to better fit with current DT initiatives by state agencies.

Second, the current literature on translation focuses on the translation of clearly-defined and legitimized ideas while overlooked how ambiguous ideas get legitimized in the field (Nielsen et al. 2013; Nielsen et al. 2020). The introduction of EA in Vietnam up until the publication of EA 1.0 provides an account of what happens when the focal innovation concept is ambiguous. In this situation, parallel sense-making and translation happened at both field- and organizational-level. The experiments allowed a collective sense-making of EA concepts, something resemble the making of an organizing vision for EA (Swanson and Ramiller 1997). What differs in this case is that the success of one pioneer was picked up by field-level organizations, allowing the EA concepts to be formalized and theorized for wider adoption.

|  | Phase 1: Experimenting (2005 – 2010) | Phase 2: Legitimizing (2010 – 2015) | Phase 3: Situated improvising (2015 – 2022) |
|---|---|---|---|
| Process | The 2005 WB Project introduced EA concepts to Vietnamese government agencies, triggering five sub-projects to adopt EA (i.e., micro translating EA concepts into organizations) | The success of pioneers inspired MIC to study success stories and came up with the first national EA framework, focusing mostly on e-government applications | MIC revised EA 1.0 into EA 2.0 and 3.0 to better fit with current DT initiatives by state agencies |
| Mechanisms | A sandbox approach in which various pioneers simultaneously adopted EA concepts; most failed and a few succeeded | A theorization process in which lessons learned from organizations (e.g., DNG) and other success stories get formalized into an EA framework | A contextualization process in which abstract concepts in EA 1.0 were expanded and contextualized in issues that are relevant to majority of the adopters |
| Impacts | Various experiments allowed pioneers to try different ways to translate EA concepts into practice as well as drew lessons on what made EA work | 19 provinces emulated DNG's approach, and the first national EA framework was published | Increase practicality and compatibility of EA concepts |
| Table 3. Case Summary | | | |

This study has certain limitations. First, we employed a case study research approach within the context of Vietnam, which may restrict generalizability to other contexts; however, it allows for the generalization of findings to theory (Yin 2009). Second, as with qualitative research methods, there is potential for bias. To mitigate this limitation, we have provided illustrative examples through quotations. In addition, we have employed a triangulation technique to validate our data, utilizing internal documents, observations, and follow-up emails. Future studies are encouraged to examine the translation process in other contexts to validate our findings here.

## Appendix A: Abbreviations

| Abbreviation | Description |
|---|---|
| AI | Artificial Intelligence |
| AITA | Authority of Information Technology Applications (AITA), now changed to Authority of Digital Transformation, is a unit under the MIC that is responsible for strategy, policy, and standards related to IT applications in state agencies. |
| CIO | Chief Information Officer |
| DICT | Department of Information and Communications Technology |
| DNG | Da Nang city |
| DT | Digital Transformation |
| EA | Enterprise Architecture |
| E-Government | Electronic Government |
| FEA | The Federal Enterprise Architecture Framework from the US Federal government |
| GSO | General Statistics Office of Vietnam |
| HAN | Ha Noi City |
| HAT | Ha Tinh Province |
| HAT | Ha Tinh province |
| HCM | Ho Chi Minh City |
| IoT | Internet of Things |
| IT/ICT | Information Technology/Information and Communication Technology |
| IS | Information Systems |
| LAO | Lao Cai province |
| MCF | Microsoft Connected Framework |
| MIC | Ministry of Information and Communications |
| ML | Machine Learning |
| OIO | The Danish Government Enterprise Architecture Framework |
| PIU | Project Implementation Unit |

| QUA | Quang Ninh province |
|---|---|
| SOA | Service-Oriented Architecture |
| TOR | Terms of Reference |
| TOGAF | The Open Group Architecture Framework |
| WB | The World Bank |

# Acknowledgments